# Low energy upscattering of UCN in interaction with the surface of solid state

A. Serebrov[1*)], A. Vassiljev[1], M. Lasakov[1], I. Krasnoschekova[1],

A. Fomin[1], P. Geltenbort[2], E. Siber[1]

[1] *Petersburg Nuclear Physics Institute, Russian Academy of Sciences, Gatchina, Leningrad District, 188300, Russia*

[2] *Institut Max von Laue – Paul Langevin, B.P.156, 38042 Grenoble Cedex 9, France*

---

[*)] *Corresponding author:* A.P. Serebrov

A.P. Serebrov

Petersburg Nuclear Physics Institute

Gatchina, Leningrad district

188300 Russia

Telephone: +7 81371 46001

Fax: +7 81371 30072

E-mail: serebrov@pnpi.spb.ru

**Abstract**

In connection with the problem of anomalous losses of ultracold neutrons (UCN) in the storage in the material traps the measurements of quasi-elastic scattering of UCN in collisions with trap surface have been carried out. It was observed that the probability of quasi-elastic scattering of UCN from energy interval 0-77 neV to 90-200 neV is $(2.2\pm0.2)\cdot10^{-8}$ per collision therefore the process of quasi-elastic scattering can not explain the effect of anomalous losses which is on the level $3\cdot10^{-5}$.

A low energy upscattering of UCN (or quasi-elastic scattering) during the reflection from liquid fomblin oil was observed at the level of $5 \cdot 10^{-6}$ per collision in a few experimental works [1-3]. The same studies demonstrated that there was a strong temperature dependence of the probability of this process. For example, in the work [2] it was shown that in the course of temperature decreasing and solidification of fomblin oil the probability of quasi-elastic scattering fell down lower then $1 \cdot 10^{-6}$ per collision. The effect of low energy upscattering in reflection from solid state has to be considerably lower. In our previous studies [2] we established only the upper limits for the probability of quasi-elastic scattering from the energy interval $0 < E < 50$ neV to the energy interval $80 < E < 150$ neV. For example, these upper limits are: $3.6 \cdot 10^{-8}$ for Cu, $3.2 \cdot 10^{-8}$ for weak-magnetic stainless steel, $3 \cdot 10^{-8}$ for graphite coating and $3 \cdot 10^{-8}$ for beryllium and beryllium coatings. As the exception to this rule the effect of magnetic quasi-elastic scattering $(4.1 \pm 0.4) \cdot 10^{-7}$ per collision was observed for a special kind of stainless steel with magnetic impurities.

However, in the course of these studies [4] we observed the effect of transmission through a perfect beryllium coating of a Si plate (wafer for electronic devices). This effect can be explained: 1) by defects of the coating, 2) by the effect of low energy heating with low probability ($\sim 2 \cdot 10^{-8}$ per collision) during the storage in the UCN trap.

To study this effect we prepared a perfect Be coating on thin glass plates (0.3 mm) without borated admixtures. The transmission factor for light for these plates was measured and was observed at the level of $10^{-9}$ or less. Nevertheless in the course of neutron measurements the transmission factor for UCN was observed again at the level of about $10^{-6}$. In order to check the hypothesis of the low energy heating with the low probability of upscattering we used the well-known scheme of the experiment [5] which is shown in Fig. 1. The time diagram of the measurement is shown in Fig. 2. UCN come from UCN turbine and fill the volume of gravitational spectrometer through the open valve 1. The valve 2 is closed at this moment. Nevertheless we can see the process of filling because a very small part of the neutron flux penetrates through the slit of the valve 2, through the glass plate 3, and reaches the UCN detector 4. Then the valve 1 is closing and the valve 2 is opening during the time interval of 120-130 s. The absorber of the gravitational spectrometer is in the position 0.75 m from the axis of the UCN guide. Therefore the absorber has to cut the UCN spectrum at the energy of 75 cm (76.8 neV) with respect to the position of the glass plate. The critical energy of glass was measured by the gravitational spectrometer and was equal to 90 cm (92.2 neV). Thus the UCN spectrum does not have to have neutrons with the energy higher then the critical energy of the glass plate. On the diagram within the time interval of 130-200 s we can see the process of cutting of the spectrum by the absorber. At the moment of 200 s the counting rate of the detector reaches the level near the background. Just at this moment the absorber is lifted to the position of 195 cm with respect to the position of the glass plate. The neutrons upscattered from the energy interval of 0-75 cm (0-76.8 neV) to the energy interval of 90-195 cm (92.2-199.9 neV) can be stored in the spectrometer trap, can penetrate through the glass plate and can be detected. On the diagram within the time interval of 200-350 s we can see the process of low energy heating and detection of these neutrons. At the moment of 350 s the valve 2 is closing for the measurements of background. We can see the excess of the counting rate with respect to background within the time interval of 200-350 s, which arises due to the low energy upscattering of UCN during the storage in the gravitational spectrometer. For the calculation of the probability of upscattering per collision we can write the following relation:

$$N = \int_0^\infty N_0 e^{-t/\tau} \alpha v \, dt = N_0 \alpha v \tau, \qquad (1)$$

where $N$ is the number of the low energy upscattered neutrons during the process of storage which is much more than the storage time $\tau$ in the spectrometer trap, $N_0$ is the number of UCN inside the spectrometer trap at the moment of 200 s, $\alpha$ is the probability of upscattering per collision, $v$ is the frequency of collision with the trap walls. Thus the simplified formula for calculation of the probability of upscattering is:

$$\alpha = N / N_0 v \tau. \qquad (2)$$

This formula does not take into account the process of the second order – like double quasi-elastic scattering because the probability of quasi-elastic scattering is much less than the probability of losses in the trap. In our experiment we can measure $N^* = \varepsilon_{up} N$ and $N_0^* = \varepsilon_{below} N_0$ instead of $N$ and $N_0$, where $\varepsilon_{up}$ is the efficiency of the detection of the upscattered neutrons, $\varepsilon_{below}$ is the efficiency of the detection of neutrons with the energy below the critical energy of glass. The efficiency of detection of upscattered neutrons depends on the transmission factor ($T$) through the glass plate, the probability of upscattered neutron to leave the spectrometer trap without losses $\tau_{st}^{up}/(\tau_{st}^{up} + \tau_{emp}^{up})$ and the efficiency of the detector for the energy interval of (90-200 neV) $E^{up}$.

$$\varepsilon^{up} = T E^{up} \tau_{st}^{up}/(\tau_{st}^{up} + \tau_{emp}^{up}), \qquad (3)$$

where $\tau_{emp}^{up}$ is the time of emptying of the spectrometer for neutrons within the energy interval of (90-200 neV).

In a similar form we can present the efficiency of detecting of neutrons with the energy below the critical energy of glass. This measurement of $N_0^*$ carried out without the glass plate for the same time interval of 200-350 s.

$$\varepsilon^{below} = E^{below}\tau_{st}^{below}/(\tau_{st}^{below} + \tau_{emp}^{below}), \qquad (4)$$

where $\tau_{emp}^{below}$ is the time of emptying of the spectrometer for neutrons with the energy below the critical energy of glass, $E^{below}$ is the efficiency of detector for the energy interval below the critical energy of glass.

In our consideration $E^{up}$ and $E^{below}$ include the transmission factor through the guide behind the glass plate. This guide is necessary to accelerate low energy UCN and improve the transmission through the aluminum window of the $^3$He detector. This guide has the coating $^{58}$NiMo with the critical velocity of 7.8 m/s to prevent the losses of neutrons for the energy interval of 90-200 neV in the course of acceleration. Our estimation shows that the ratio $E^{below}/E^{up}$ is about one with the accuracy of 20-30%. In this approximation we can write the following formula for calculation of the probability of upscattering from the energy interval of 0-77 neV to the energy interval of 90-200 neV:

$$\alpha = \frac{1}{\nu\tau_{st}^{below}T} \cdot \frac{N^*}{N_0^*} \cdot \frac{\tau_{st}^{below}}{\tau_{st}^{up}} \cdot \frac{(\tau_{st}^{up} + \tau_{emp}^{up})}{(\tau_{st}^{below} + \tau_{emp}^{below})}. \qquad (5)$$

All parameters in this formula besides $\nu$ were measured using the gravitational spectrometer which allows to select the necessary part of the spectrum. The measurements were done with the glass plate and without it. The average value of the frequency of collisions was obtained by means of Monte Carlo calculations.

Below are presented the necessary values for the calculation of the probability of upscattering $\alpha$: $N_0^*=2.06 \cdot 10^5$; $N^*=1.03\pm0.08$; $T=0.253\pm0.008$; $\tau_{st}^{below}=(155.0\pm0.6)$ s; $\tau_{st}^{up}=(79.7\pm0.2)$ s; $\tau_{emp}^{below}=(37.9\pm0.2)$ s; $\tau_{emp}^{up}=(39.9\pm0.4)$ s; $\nu=7$ Hz. One can see that only one upscattered neutron was detected out of $2.06 \cdot 10^5$ neutrons which had about $10^3$ collisions during the time interval of 150 s. The accuracy of this calculation of $\alpha$ is mainly determined by the statistical accuracy of the number of upscattered neutrons per one experimental run ($N^*=1.03\pm0.08$).

$\alpha=(2.2\pm0.2) \cdot 10^{-8}$ per collision  (6)

However, due to the fact that the spectrum of upscattered neutrons was not measured, a deviation of the shown value $\alpha$ is possible by the factor of about 2.

In accordance with conditions of the experiment the value $(2.2\pm0.2) \cdot 10^{-8}$ corresponds to the probability of quasi-elastic scattering of UCN from the energy interval of 0-77 neV to 90-200 neV. This probability of upscattering is less by the factor of about 30 than the probability of upscattering obtained in the work [6] for beryllium.

The reasons of this discrepancy are not quite clear. They can be connected:

1) with the method of calculation of $\alpha$ in the work [6] (unfortunately there is no data in the work [6] to compare some details of the methods of calculation);
2) with the conditions the experiment conduction in the work [6]. For example, the dust on the surface of samples can produce enhanced upscattering process as it was shown in the works of the same experimental group [7]. A strong heating of samples and the spectrometer was used in the work [6] to remove organic from the surface. It could produce some number of fish scales on the surface which can increase the upscattering due to its own vibration.

From the data of our experiment we have to conclude that the probability of quasi-elastic scattering with the energy transfer of about 100 neV is about $(2.2\pm0.2)\cdot10^{-8}$ per collision for the surface of Be coating on the Cu trap. The probability of anomalous losses [8] was $3\cdot10^{-5}$ per collision for the similar coating of the trap and for the solid Be trap also. Therefore the process of quasi-elastic scattering ($\Delta E=100$ neV) can not explain the effect of anomalous losses as it was proposed in [9]. The measurement of the probability of quasi-elastic scattering in the work [6] has been overestimated due to methodical reasons by the order of magnitude or more.

The work has been supported by the Russian Foundation of Basic Research (Grant No 04-02-17440).

**Figure captions**

Fig. 1. The scheme of the experiment.

Fig. 2. The time diagram of the measurement.

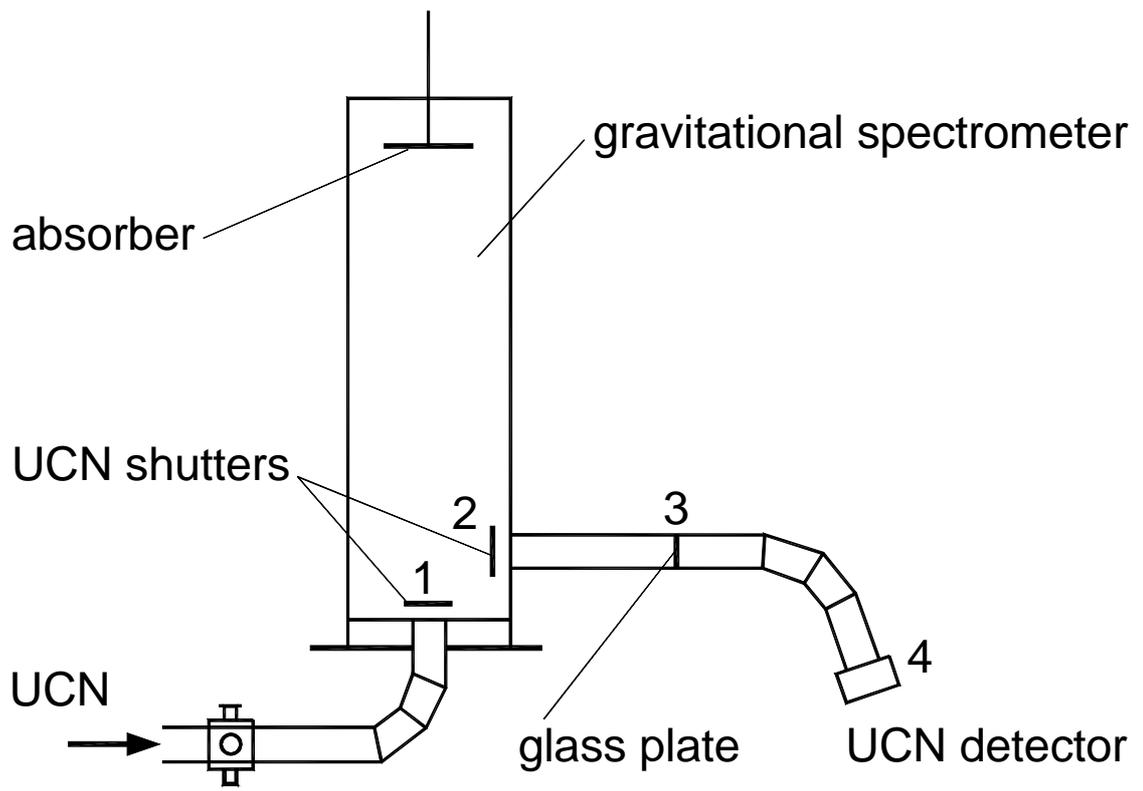

Fig. 1.

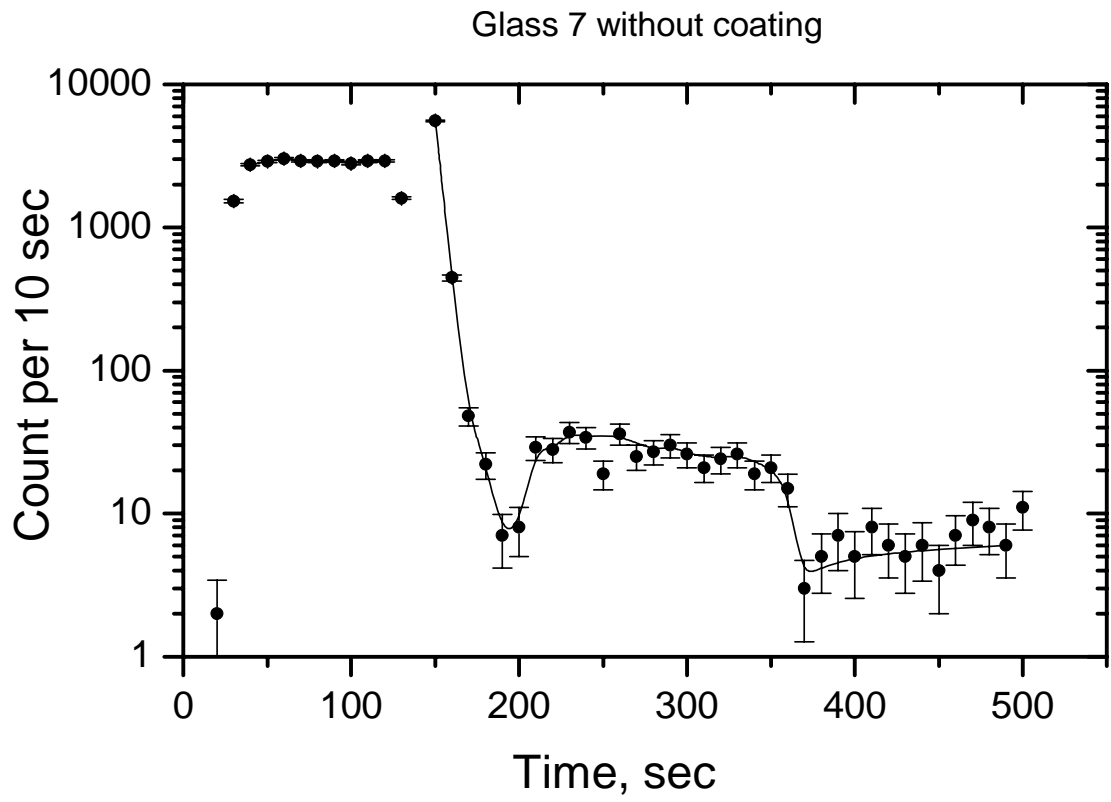

Fig. 2.